\documentclass[unnumsec,webpdf,contemporary,large]{oup-authoring-template}%

\graphicspath{{Fig/}}

\theoremstyle{thmstyleone}%
\theoremstyle{thmstyletwo}%
\theoremstyle{thmstylethree}%

\begin{document}

\journaltitle{Bioinformatics}
\copyrightyear{2026}
\pubyear{2026}
\appnotes{Application Note}

\firstpage{1}


\title[Regression and network estimation in QIIME~2]{Sparse regression, classification, and microbial network estimation in QIIME~2 with q2-classo and q2-gglasso}

\author[1,2,$\ast$]{Oleg Vlasovets}
\author[3]{Fabian Schaipp}
\author[4]{L\'eo Simpson}
\author[5]{Evan Bolyen}
\author[5]{J. Gregory Caporaso}
\author[1,2,6]{Christian L. M\"uller}

\authormark{Vlasovets \textit{et al.}}

\address[1]{\orgdiv{Computational Health Center}, \orgname{Helmholtz Munich}, \orgaddress{\state{Neuherberg}, \country{Germany}}}
\address[2]{\orgdiv{Ludwig-Maximilians-Universit\"at M\"unchen}, \orgname{Department/Faculty}, \orgaddress{\state{M\"unchen}, \country{Germany}}}
\address[3]{\orgdiv{Technical University of Munich}, \orgname{Department/School}, \orgaddress{\state{M\"unchen}, \country{Germany}}}
\address[4]{\orgdiv{Albert-Ludwigs-Universit\"at Freiburg}, \orgname{Department/Faculty}, \orgaddress{\state{Freiburg}, \country{Germany}}}
\address[5]{\orgdiv{Pathogen and Microbiome Institute}, \orgname{Northern Arizona University}, \orgaddress{\state{Flagstaff, AZ}, \country{USA}}}
\address[6]{\orgdiv{Center for Computational Mathematics}, \orgname{Flatiron Institute}, \orgaddress{\state{New York, NY}, \country{USA}}}

\corresp[$\ast$]{Corresponding author. \href{mailto:oleg.vlasovets@helmholtz-munich.de}{oleg.vlasovets@helmholtz-munich.de}}




\abstract{
\textbf{Motivation:} Statistical analysis of microbial count data derived from 16S rRNA or metagenomics sequencing poses unique challenges due to the sparse, compositional, and high-dimensional nature of the data. While QIIME~2 already provides many tools for data pre-processing and analysis, plugins for statistical regression, classification, and microbial network estimation tailored to compositional count data are relatively scarce.\\
\textbf{Results:} We present \texttt{q2-classo} and \texttt{q2-gglasso}, two novel QIIME~2 plugins that implement penalized regression, classification, and graphical modeling approaches for microbial compositional data. \texttt{q2-classo} enables the prediction of a continuous or binary outcome of interest using compositional microbiome data as predictors. Both sparse log-contrast regression and classification, as well as tree-aggregated log-contrast models are available. \texttt{q2-gglasso} enables the estimation of taxon-taxon association networks through sparse graphical model estimation, such as, e.g., the SPIEC-EASI framework, as well as adaptive and latent graphical models. The latent model can decompose taxon-taxon associations into a sparse direct interaction matrix and a latent (low-rank) matrix which enables robust principal component embedding of a data set. Within the QIIME~2 ecosystem we demonstrate their application  on the Atacama soil microbiome dataset, illustrating robust model selection, classification, and microbial network estimation with covariates and latent factors.\\
\textbf{Availability:} The software is freely available under the BSD-3-Clause License. Source code is available at \url{https://github.com/bio-datascience/q2-gglasso} and \url{https://github.com/bio-datascience/q2-classo-latest}, with installation through QIIME~2 and Docker.\\
\textbf{Contact:} \href{mailto:oleg.vlasovets@helmholtz-munich.de}{oleg.vlasovets@helmholtz-munich.de}\\
}
\keywords{QIIME~2, high-dimensional statistics, compositional data analysis, microbial networks, sparse regression, graphical lasso}


\maketitle

\section{Introduction}
Microbial abundance data, derived from high-throughput amplicon or metagenomics sequencing experiments and subsequent 
dedicated pre-processing \cite{callahan2016dada2}, comprise microbial counts in form of amplicon sequence variants (ASVs), 
operational taxonomic units (OTUs), or metagenomics-derived OTUs (mOTUs). These counts are characterized by an excess number 
of zeros and carry only compositional (or relative abundance) information \cite{gloor2017microbiome}. Moreover, in a typical 
microbiome dataset the number of microbial features is larger than the number of available samples (i.e., data are in the ``high-dimensional" regime), making downstream 
statistical estimation tasks, such as differential abundance testing \cite{morton2019establishing}, regression \cite{lin2014variable}, or covariance and network estimation \cite{kurtz2015sparse}, challenging. 

\begin{figure*}[!t]
\centering
\includegraphics[width=\textwidth]{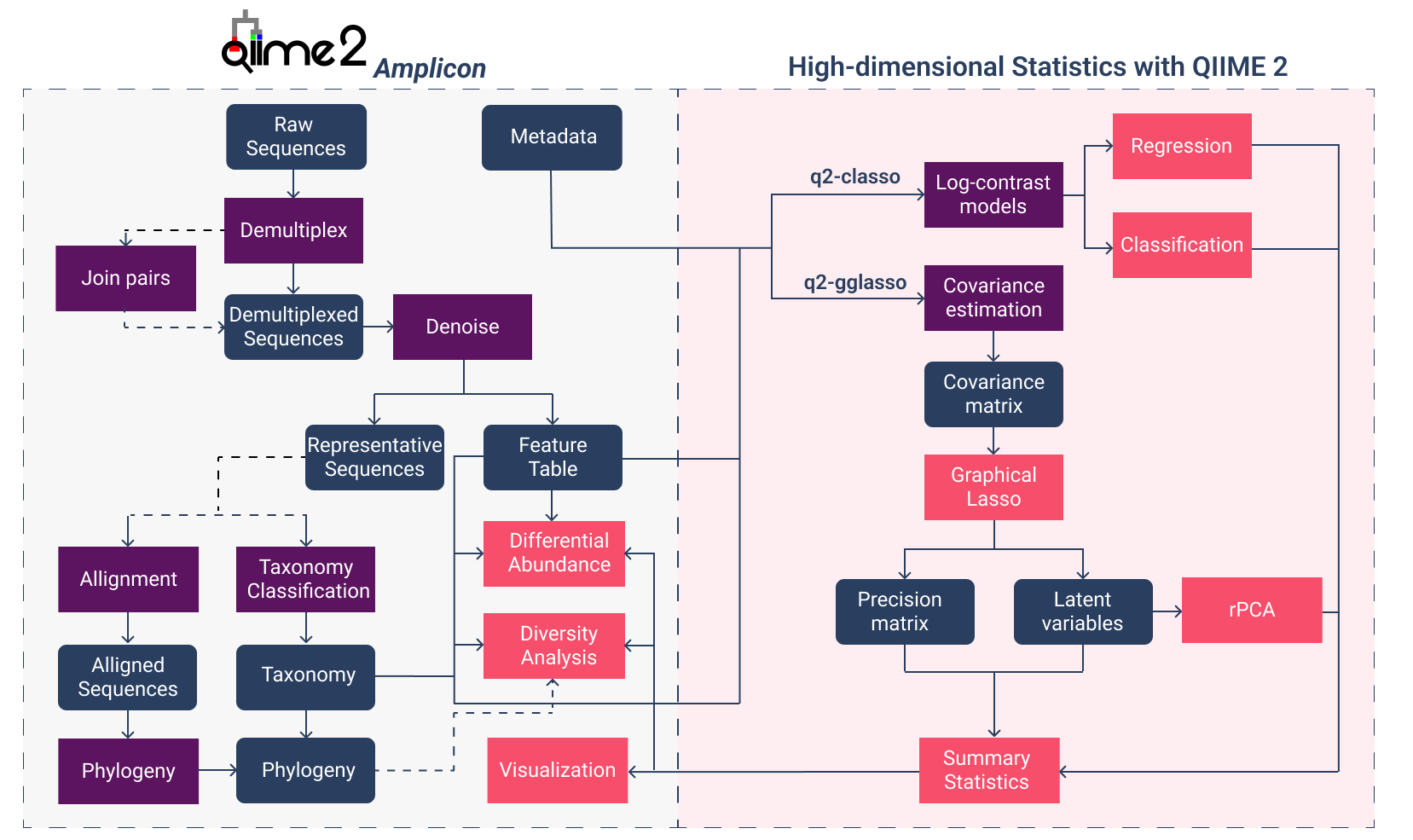}
\caption{QIIME~2 workflow and its high-dimensional statistics extension. The left panel illustrates typical amplicon processing steps in QIIME~2. Feature Table, Taxonomy objects and available Metadata serve as input to the high-dimensional statistics plugins \texttt{q2-classo} and \texttt{q2-gglasso} (right).}
\label{fig:q2_stats}
\end{figure*}

In this contribution, we introduce \texttt{q2-classo} and \texttt{q2-gglasso}, two novel QIIME~2 plugins, that 
implement principled methods for (high-dimensional) statistical regression, classification, and network estimation for 
microbial compositional data. This enables microbiome researchers easy access to these tools within the QIIME~2 ecosystem, an open-source microbiome bioinformatics platform, recognized for its robust functionality, reproducibility through automatically tracked data provenance, extensibility through a module-based plugin architecture, and broad user and developer community \cite{bolyen2019reproducible}. 
While QIIME~2 already comprises a number of plugins for pre-processing raw sequences 
\cite{callahan2016dada2}, diversity analysis \cite{bolyen2019reproducible, raspet2025facilitating}, community 
structure investigation \cite{bolyen2019reproducible, lozupone2011unifrac, mcdonald2018striped}, and differential 
abundance testing \cite{fernandes2013anova, lin2020analysis}, our plugins complement existing statistical approaches 
for regression, classification, and network estimation (see Fig.~\ref{fig:q2_stats}).
For instance, the popular \texttt{q2-sample-classifier} \cite{bokulich2018q2} plugin offers standard machine learning 
methods such as Ridge and Lasso regression, but does not provide log-contrast regression models tailored to 
compositional data. \texttt{q2-longitudinal}~\cite{bokulich2018q2} supports linear mixed-effects models and paired-
sample comparisons for low-dimensional time-series. \texttt{songbird}~\cite{morton2019establishing} uses multinomial 
regression specifically for differential abundance testing. \texttt{Qurro} \cite{fedarko2020visualizing} enables the 
visualization of quantitative ranks and ratios, often in the form of log-contrasts or log-ratios, but cannot explicitly 
build combinations of log-ratios in a data-driven manner. In the context of microbial network estimation, the currently 
only available tool is \texttt{q2-SCNIC}~\cite{shaffer2023scnic}. SCNIC (Sparse Correlation Network Investigation for Compositional data) constructs standard pairwise correlation networks from feature tables using metrics such as SparCC \cite{friedman2012inferring}, detects co-occurrence modules, and allows for visualization of correlated features and modules. 

We next describe the key objectives of the two plugins, outline their mathematical underpinning, and their embedding within QIIME~2. We also illustrate a practical use case on the Atacama soil microbiome dataset.  

\section{Methods and implementation}

\subsection{Log-contrast regression and classification for microbial relative abundances with \texttt{q2-classo}}
\texttt{q2-classo} enables the prediction of a continuous or binary outcome of interest from high-dimensional microbial abundance data. The underlying statistical model is the so-called log-contrast model, as introduced in \cite{aitchison1984log} for regression, where an outcome of interest, e.g., an environmental covariate or the disease status of patient, is implicitly modeled as a linear combination of log-ratios of individual compositional parts (e.g., microbial relative abundances). This model has been extended to sparse (penalized) estimation in \cite{lin2014variable,shi2016regression} when more features (taxa) are available than samples. Since then, the model and its underlying statistical estimation has been further refined to allow for robust regression \cite{combettes2021regression, mishra2022robust}, joint estimation of noise \cite{combettes2021regression}, taxonomic tree-aggregated regression (\texttt{trac} \cite{bien2021tree}), and classification \cite{Simpson2021}. \texttt{q2-classo} inherits these log-contrast models from the \texttt{c-lasso} package \cite{Simpson2021} and makes them available in QIIME~2. 

Figure~\ref{fig:q2_stats} shows a typical QIIME~2 amplicon sequencing data analysis pipeline and its connection to 
\texttt{q2-classo}. While a typical analysis comprises multiple functionalities, the QIIME~2 artifacts \texttt{Feature Table[Frequency]}  and \texttt{FeatureData[Taxonomy]}, as well as Metadata (containing outcomes of interests) serve as direct inputs to \texttt{q2-classo}. Prior to model fitting, \texttt{q2-classo} applies a centered log-ratio transform  to the feature counts to account for the compositional nature of the data. Broadly, three model classes are available: (i)~sparse log-contrast regression for continuous outcomes, (ii)~sparse log-contrast 
classification for binary outcomes, and (iii)~tree-aggregated log-contrast (\texttt{trac}) models which can incorporate 
taxonomic information. To guard against overfitting, \texttt{q2-classo} provides three model selection strategies: a 
theoretically-derived fixed penalty \cite{shi2016regression, combettes2021regression}, $k$-fold cross-validation, and 
stability selection \cite{meinshausen2010stability, combettes2021regression, jiang2022utilizing}. After model estimation, \texttt{q2-classo} stores the model output, including selected taxa, their effect sizes, and model selection curves, in QIIME~2 \texttt{.qza/.qzv} files, which also enable visualizations for interactive inspection.

\subsection{Microbial network estimation with \texttt{q2-gglasso}}
\texttt{q2-gglasso} enables the estimation of taxon-taxon association networks via the application of so-called 
graphical modeling techniques. The principle idea is to estimate a \emph{sparse} inverse covariance (or \emph{precision}) matrix 
from a given sample covariance matrix of the (microbial) features using a penalized likelihood \cite{dempster1972covariance}, also referred to as graphical lasso \cite{friedman2008sparse}. The non-zero coefficients of the estimated inverse covariance matrix are related to partial correlations and can be interpreted as associations between the different features (taxa) and summarized as network. Important extensions of 
this model include the estimation of joint graphical models across multiple data sets \cite{danaher2014joint} and latent 
graphical models \cite{chandrasekaran2010latent}. The latter approach estimates both a sparse inverse covariance and a dense latent (low-rank) matrix that can adjust for hidden confounders and biases. The low-rank matrix, in turn, can then be used for robust principal component analysis (rPCA), similar in spirit to \texttt{q2-deicode} \cite{martino2019novel}. In the microbiome context, graphical modeling has been used in SPIEC-EASI (SParse InversE Covariance Estimation for Ecological Association Inference) \cite{kurtz2015sparse,kurtz2019disentangling}, gCODA \cite{fang2017gcoda}, and others (see e.g., \cite{peschel2021netcomi} for an overview). \texttt{q2-gglasso} inherits these network estimation schemes through the \texttt{GGLasso} package \cite{Schaipp2021}, an efficient Python implementation of a wide range of sparse graphical models, and makes them available in the QIIME~2 ecosystem.     

As shown in Fig.~\ref{fig:q2_stats}, \texttt{q2-gglasso} takes as primary input \texttt{Feature Table[Frequency]} and estimates 
taxon-taxon association networks via graphical modeling techniques. To address the compositional nature of the data, the 
centered log-ratio (clr) or a modified clr (mclr) transformation \cite{yoon2019microbial} is applied prior to covariance estimation. The following major graphical model classes are available:
(i)~standard graphical lasso \cite{friedman2008sparse}, which recovers a sparse precision matrix representing conditional dependence structure among taxa — analogous to the SPIEC-EASI framework \cite{kurtz2015sparse}; (ii)~multimodal graphical lasso, which enables combined estimation of taxa–covariate associations (through adaptive penalization of the different modalities); and (iii)~a sparse+low-rank model \cite{chandrasekaran2010latent,kurtz2019disentangling} that decomposes the precision matrix into a sparse direct interaction component and a low-rank latent factor component, enabling latent variable correction and robust PCA of the microbiome data \cite{kurtz2019disentangling}. After model estimation, \texttt{q2-gglasso} stores the model output, including the estimated precision matrix, network summary statistics, low-rank components, and interactive network and heatmap visualizations, in QIIME~2 \texttt{.qza/.qzv} files for further visualizations and interactive inspection.

\section{Use case on Atacama soil dataset}
We illustrate the functionality of both plugins by focusing on a subset 13 highly abundant taxa across 50 samples of the Atacama soil microbiome dataset \cite{neilson2017significant}.
To illustrate log-contrast regression with \texttt{q2-classo}, we predict average soil temperature as a continuous outcome. For classification, we considered vegetation type as a binary outcome. For both prediction tasks, stability selection consistently identified a small set of predictive taxa for the respective outcome.  

To illustrate network estimation with \texttt{q2-gglasso}, we used the same subset of data to 
first estimate a sparse inverse covariance matrix to identify the partial correlation structure 
among microbial taxa. We showcase how to include the available covariate information and learn 
joint partial correlations over taxa \emph{and} environmental covariates. Finally, we also 
learned a latent graphical model on the taxon data, yielding a sparse and a low-rank component. 
We finally show that the \emph{inferred} latent components from the taxon data alone correlate well with the key environmental covariates. All estimated networks are visualized as interactive heatmaps within the QIIME~2 framework. 

The detailed tutorial and reproducible pipeline examples are available in the online 
documentation at \url{https://vlasovets.github.io/q2-hdstats-docs/intro.html}.
\section{Availability and Implementation}
Both plugins are Python implementations integrated into QIIME~2.  

\noindent Software availability:
\begin{itemize}
    \item q2-gglasso: \href{https://library.qiime2.org/plugins/bio-datascience/q2-gglasso}{library.qiime2.org/.../q2-gglasso}
    \item q2-classo: \href{https://library.qiime2.org/plugins/bio-datascience/q2-classo-latest}{library.qiime2.org/.../q2-classo-latest}
    \item License: BSD-3-Clause
\end{itemize}

Data availability: The Atacama soil microbiome dataset \cite{neilson2017significant} is available from the European Nucleotide Archive (ERP019482). Processed count tables, covariates, and SILVA-based taxonomic annotations are also provided.

\section{Conclusion}
\texttt{q2-classo} and \texttt{q2-gglasso} provide capabilities for compositionally-aware 
sparse regression, classification, and microbial network estimation directly within 
QIIME~2. Making these state-of-the-art high-dimensional statistics tools available in the 
QIIME~2 ecosystem allows, in turn, easy integration with existing analysis workflows, automated provenance tracking ensuring transparency and reproducibility, and accessibility to a broad user base.

\section*{Funding}
O.V. is supported by the Helmholtz Association through the joint research school Munich School for Data Science (MUDS).
E.B. and J.G.C. were funded in part by NIH National Cancer Institute Informatics Technology for Cancer Research Award 1U24CA248454-01 to JGC.

\section*{Conflict of Interest}
The authors declare no conflict of interest.

\section*{Author contributions statement}
O.V. developed and tested \texttt{q2-gglasso}. F.S. developed the \texttt{GGLasso} Python 
library which \texttt{q2-gglasso} is built on. E.B., L.S., and O.V. developed and tested 
\texttt{q2-classo}.  O.V. conducted all formal analyses and wrote the original draft. 
C.L.M. contributed to conceptualization, methodology, editing the draft and supervision. 
J.G.C. provided resources and infrastructure support. All authors reviewed the manuscript.

\section*{Acknowledgments}
This work is supported in part by funds from the Helmholtz Association through the joint research school Munich School for Data Science (MUDS).

\bibliographystyle{plain}
\bibliography{reference}
\end{document}